# A low-cost flexible instrument made of off-the-shelf components for pulsed eddy current testing: overview and application to pseudo-noise excitation.

Hamed Malekmohammadi[1*], Andrea Migali[2], Stefano Laureti[2], Marco Ricci[2]

[1]*Department of Engineering, Polo Scientifico Didattico di Terni,*
*University of Perugia, Strada di Pentima 4, 05100 Terni, Italy*
hamed.malekmohammadi@unipg.it

[2]*Department of Informatics, Modeling, Electronics and Systems Engineering,*
*University of Calabria, Via Pietro Bucci, 87036 Arcavacata, Rende CS, Italy*
{marco.ricci, stefano.laureti}@unical.it, mglndr97a05c352w@studenti.unical.it

[*] *Correspoding Author*

*Abstract:*

**A flexible and low-cost device for eddy current non-destructive testing made of off-the-shelf components is described. The proposed system is compact and easy to operate, and it consists of a dual H-bridge stepper motor driver, a coil winded in-house on an additively manufactured support, a tunnel magnetoresistance sensor, and a data generation/acquisition module. For the latter, two different commercial devices have been used, and both setups have been then tested on a benchmark sample to detect small artificial cracks. The system can flexibly generate square pulse or square wave with tunable duration and frequency, as well as pseudo-noise binary waveforms that are here used in combination with pulse-compression to increase the inspection sensitivity with respect to standard pulsed eddy current testing. A benchmark sample was analysed, and all the defects were correctly located, demonstrating the good detection capability of the sensor. This was achieved by assembling a very low-cost handy device, which can be further improved in portability and performances with the use of different off-the-shelf components, and that can be easily integrated with single-board PC, paving the way for future developments in this field.**

I. INTRODUCTION

Non-destructive testing (NDT) applications are fast-expanding, and this requires efficient solutions to be used in real applications. With the development of the 'Internet of Things' and the 'Industry 4.0' paradigms, NDT measurements are expected to be increasingly integrated into production lines, not only as a mean to detect possible defects but also to monitor the product quality and providing instantaneous feedback of the process itself. Similarly, new structural health monitoring procedures can be developed by using various sensor types and NDT techniques such as eddy current testing (ECT).

In this framework, a key goal is realising a low-cost and handy instrumentation, made of off-the-shelf components that can provide high sensitivity and flexibility at the same time. To accomplish this aim, we realized a hardware and software system for pulsed eddy current (PEC) testing relying on: (1) a custom sensor consisting of a coil for generating the excitation magnetic field and a tunnelling magnetoresistance (TMR) sensor to measure the induced magnetic field; (2) a low-cost electronic board containing a driver for DC stepping motors, which feeds the coil; (3) the use of binary pseudo-noise (PN) waveforms that allow exploiting various advanced signal processing strategies such as pulse-compression (PuC).

PEC is nowadays widely used in commercial instruments for various types of inspections such as detecting corrosion under insulation, sub-surface defects, pipe-wall thinning, and insulating/coating material thickness measurement [1, 2, 3, 4] and it still attracts many research efforts to further improve the technique. New theoretical aspects, analytical and numerical simulation tools, tailored processing techniques and applications are being investigated [5, 6, 7, 8].

The main pros of PEC are nowadays well acknowledged in ECT community. A pulsed signal allows a continuous range of frequencies to be excited, and even a very short pulse contains low-frequency components capable of penetrating deeply into conductive materials. This feature lies at the heart of the PEC applications mentioned above. At the same time, PEC signals can be processed in both time and frequency domains, and it was found that time-domain analysis is very powerful. Features such as lift-off invariant (LOI) point, time-phase, and the long-term behaviour exploited in wall thinning analysis have been reported in the literature [9, 10, 11, 12].

From a theoretical point of view, if the excitation pulse is short enough, a PEC output data can be interpreted as the estimate of the impulse response of the sample in the measurement point. In practice, in order to deliver enough energy to the system, usually rectangular pulses, square wave bursts or long pulses equivalent to a step excitation are used. In the latter case, the impulse response can be obtained by calculating the derivative of the step response, but the signal-to-noise ratio (SNR) degrades at high frequencies, while for the other cases the finite duration of the input pulse must be considered. A deconvolution process could be used to retrieve the impulse response, but this would introduce unwanted mathematical noise, so the sample response to finite duration pulse is directly processed.

There is not a preferred approach to follow, as one can shift among the various possibilities by only changing the pulse duration or by repeating the pulse periodically. So, without losing generality, hereafter we consider as the standard PEC input signal a rectangular pulse of duration $T$ and amplitude $A$, as for Eq. (1) where $\theta(t)$ is the Heaviside step function.

$$\Pi(t,T) = A\big(\theta(t) - \theta(t-T)\big) \qquad (1)$$

The excited bandwidth ($BW$) goes from DC to a cut-off frequency, which is related to the selected duration by $BW \cong \frac{1}{2T}$, so one must tailor the pulse duration by considering any a priori information about the sample to be inspected and the defects of interest.

What is important from an experimental point of view is the capability of driving the excitation coil with the desired waveform (rectangular pulse, square wave, step) so as to control the eddy current generation inside the sample and guaranteeing the fast switching of the current from an amplitude level to the other. As it will be shown in Section II, for the present application both the requirements for the excitation signal can be effectively ensured by using DC or stepping motor drivers.

It is worth to mention that the proposed instrument can be easily replicated using a variety of similar components: many possible drivers can be found in the market that vary in terms of switching time, maximum current and other characteristics. Also, the excitation coil and the receiving sensor must be designed or selected for a specific problem.

The aim of this paper is not presenting a universal optimal sensor but indeed demonstrating how cheap and off-the-shelf hardware components and signal processing can be combined to provide an effective flexible instrument.

For this purpose, we use the binary PN sequences as another, and quite powerful, possible form of excitation for PEC, *i.e.* the PN-PEC. Such signals have similar characteristics in terms of spectral content and hardware requirements of standard PEC, but they can improve the inspection capabilities of the system when combined with proper processing.

The paper is organized as follows: the hardware system parts are detailed in Section II while in Section III the theory of pseudo-noise PEC (PN-PEC) and the related post-processing are described. Section IV elaborates on experimental results obtained with PN-PEC. Conclusions and future work are drawn in Section V.

## II. Hardware system

### II.1. Eddy Current custom probe

The probe consists of two mirrored D-shaped coils, with independent excitation, and a TMR sensor placed in the middle point of the two D's, see Figure 1. This scheme follows and improves the probes described in [13, 14, 15, 16], which demonstrated good sensitivity in terms of defect detection and localisation. The independent excitation of two coils, here implemented, allows two main working modalities to be easily selected by external electronics or by signal design: one generating the primary magnetic field parallel to sample surface in the space between two coils, the other with the magnetic field perpendicular to the sample surface. Hereafter this probe will be referred as ϕ-probe, to distinguish it from other recent ECT probes using a pair of D-shaped coils, both for the excitation [17, 18] or as pick-up one [19], and from the Θ coil [20]. The coil support has been designed in CAD software and then realized using additive manufacturing technology via a commercial Ultimaker 3 Extended 3D printer (Ultimaker, Geldermalsen, Netherlands).

The TMR sensor used in this work was a TMR2905 from MultiDimension Technology®

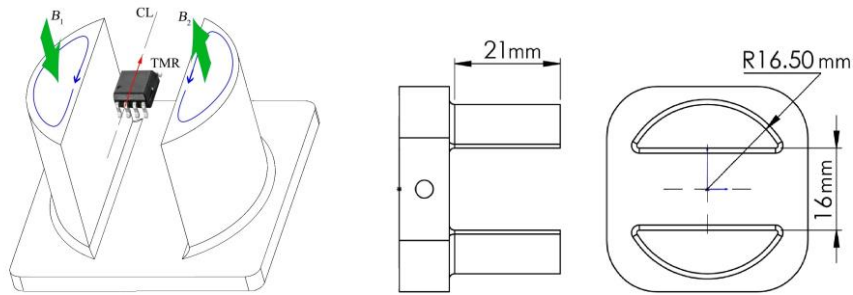

Figure 1. The sketch of the EC coil (right) and the probe configuration (left)

(Zhangjiagang, Jiangsu Province, China) having a very high sensitivity (50-60 mV/V/Gauss) and less hysteresis than giant magnetoresistive (GMR) sensors. Figure 2 shows the response of the TMR2905 to an applied magnetic field in the range of ±15 Oe and ±30 Oe when the TMR is biased at 1 V [21]. The pin-out is the same as Hall sensors, and they also have bipolar sensitivity without the need for a bias magnetic field, neither of any conditioning. All these features make TMR extremely attractive for ECT applications.

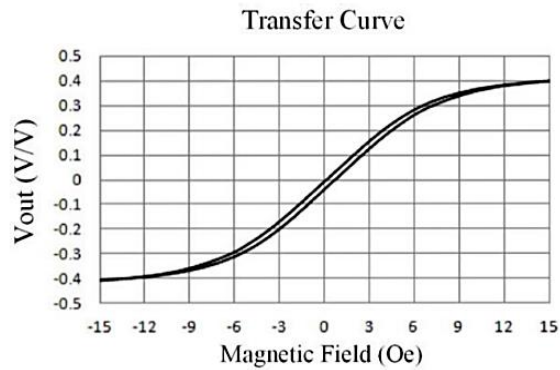

Figure 2. Response of the TMR sensor to the applied magnetic field [21]

As mentioned above, the TMR sensor is placed at the centre of the two-coil system, with the axis of sensitivity aligned parallel to the straight side of the D's. By this way, either in the case where the two coils are excited with the same current orientation (*e.g.* clockwise) or with opposite current fluxes (*e.g.* one clockwise, the other counter-clockwise), the sensor should theoretically measure a very low magnetic field component over a sound point. This is to avoid saturation of the sensor, the onset of any nonlinear phenomena by working in the linear region, and to reducing hysteresis. As an example, in Figure 1 is reported the configuration in which the magnetic field generated is parallel to the sample surface, i.e. it goes from the straight side of one coil to the straight side of the other one, and it comes from one coil to the other. The TMR is placed on the centreline of two coils and its sensitivity axis (red arrow), is perpendicular to the B-field direction.

### II.2. Current Pulser

The coils are excited using an L298N dual H-Bridge motor driver integrated with a single board module containing the ancillary circuitry for the control digital lines and the supply. The schematic diagram of the module is shown in Figure 3. The cost of the module is around one euro to a few euros, and many different modules are available containing the L298N driver, including reduced size versions using integrated circuit packages for the L298N.

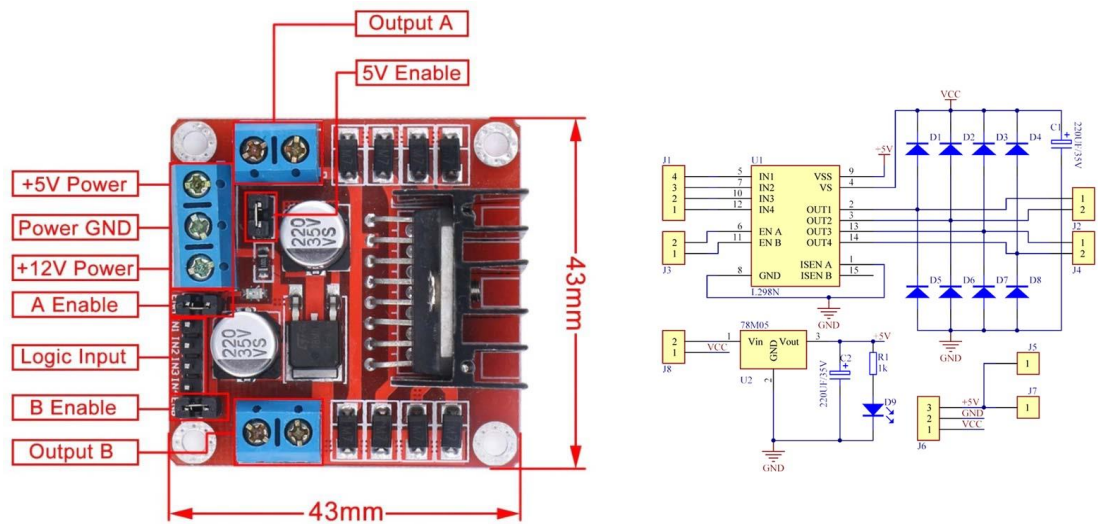

**Figure 3. Schematic diagram of the L298N motor driver card**

The module has two outputs that allow for driving two coils independently, so as change the resulting magnetic field lines in the proximity of the TMR sensor, from being perpendicular to parallel with respect to the SUT's surface, as mentioned above. It is also possible to supply one coil at the time, but this option was not explored in this paper.

The L298N component is a dual H-bridge power driver made specifically for driving inductive loads. From the diagram depicted in Figure 3 it is possible to track the function performed by each signal:

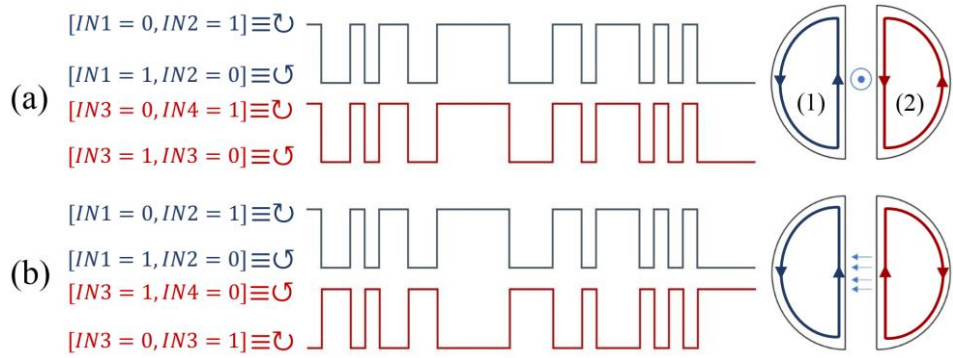

**Figure 4.** Excitation cases and generated fields: (a) in-phase currents and perpendicular magnetic field, (b) out-of-phase currents and tangential magnetic field.

- the two inductive loads, *i.e.* the coils, are connected between OUT1 and OUT2 and between OUT3 and OUT4, respectively;
- the enable signals EN_A and EN_B are connected directly to the four AND gates, whose task is to generate the control signals of two H-Bridge transistors. When the two enables are at a low state, the AND gates produce a low output making the transistors not operating. In this condition, it is not possible to power the load, regardless of the input signals. On the contrary, is possible to drive the inductive loads through the input lines when the enables are at a high state;
- IN1 and IN2 are the control signals of the first H-Bridge. IN1 directly reaches the input from the AND1 gate, while its negated counterpart reaches the AND2 logic gate. The same configuration is adopted for IN2 and the AND3 and AND4 gates.

Consequently, the following states are obtained when *e.g.*, EN_A = 1:

The two configurations $[IN1 = 0, IN2 = 1] \equiv \circlearrowleft$ and $[IN1 = 1, IN2 = 0] \equiv \circlearrowright$ are characterized by currents flowing in opposite direction within the load. The same behaviour holds for the second H-Bridge, controlled by EN_B, IN3 and IN4.

A schematic representation of these two excitation cases is depicted in Figure 4. In the case (a), both coils are fed with same binary sequence that swaps the current flowing from clockwise to counter-clockwise, but the currents in the straight edges of both the coils have always opposite direction. The dominant field lines in the space between the two coils are perpendicular to the SUT surface (*z*-component). In the case (b), the control of the output 2 is the complementary of that of output 1. The current in the straight edges of the coils flows always parallel, therefore the tangential field lines are dominant at the TMR location.

**Table 1. Logic function of the input signals**

| IN1 | IN2 | AND1 | AND2 | AND3 | AND4 | T1 | T2 | T3 | T4 |
|-----|-----|------|------|------|------|----|----|----|----|
| 0   | 1   | 0    | 1    | 1    | 0    | 0  | 1  | 1  | 0  |
| 1   | 0   | 1    | 0    | 0    | 1    | 1  | 0  | 0  | 1  |

It is worth to note that the system can easily generate a unipolar square pulse of current/magnetic field, with arbitrary current direction and duration, so as allowing the choice of the most suitable excitation for the PEC application involved.

To check the characteristics of the pulser, three single current pulses with different durations were generated in the coil (1) and the magnetic field in its proximity was measured using the TMR sensor. This is an indirect measure of the excitation current waveform provided by the L298. Figure 5(a-c) depict the TMR output voltage measured for a square pulse having a time duration of 1, 0.1 and 0.01 ms respectively.

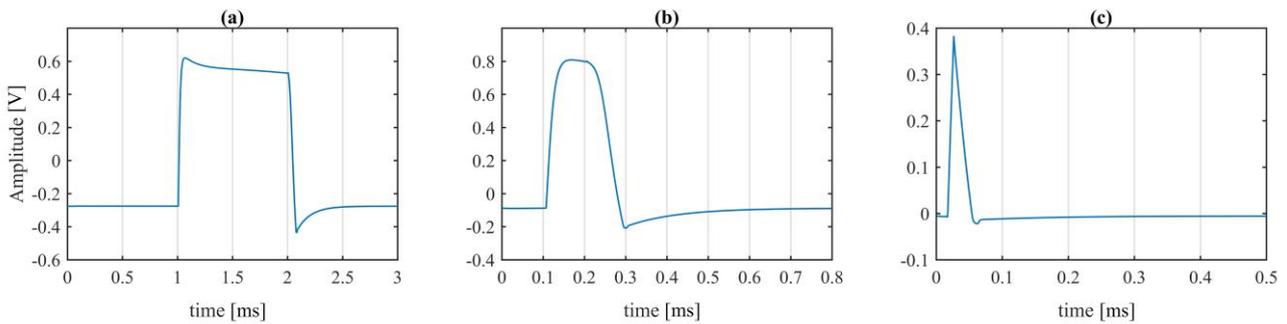

Figure 5. Example of responses of the system to various signals: (a) to (c) single pulse with 1 ms, 0.1 ms and 0.01 ms respectively.

It can be noticed that the system performs well in the case of longer pulses and poorly for short pulse of 0.01 ms, mainly for the long fall time, which is more than double the rise time. Nonetheless, this difference is due to the output enable switching from ON to OFF.

In the case of square waves or binary sequences, the inversion of the current flow does not require commuting the enable signal. Further, for the benchmark here tested, with subsurface inner defects at depths also beyond 2 mm, the pulser speed is enough to excite the bandwidth of interest, up to 10-20 kHz. In the case higher frequencies should be excited, one can chose modules containing drivers with faster switching times such as the A4988 from Allegro MicroSystems or the L6208 from ST Microelectronics.

### II.3. Measurement Setup 1

The Setup 1 is shown in Figure 6. A DC power supply was used to power both the L298N card and the TMR2905 sensor, which was fed with 5 V. The L298N card was connected to the digital lines of a National Instruments™ NI-USB 6361 Multifunction I/O device: two digital I/O lines were used to control the enable signals EN_A and EN_B of the L298N, while four other ones were instead used as input signals IN1-4 for the two H-bridge. The Output A and Output B of the driver where connected to the two coils of the ϕ-probe. The same NI-USB 6361 was exploited for acquiring the output voltage signal from the TMR through one input channel of the analog-to-digital-converter (ADC) with 16-bit of resolution.

An *x-y-z* scanning system was used to scan over a regular grid of points and the TMR output signal was saved for every $(x_k, y_l, z_m)$ measurement point as a MATLAB® file for further processing. The scan process, the signal generation and acquisition, and the input variables were managed by a virtual instrument (VI) developed in LabVIEW™ environment.

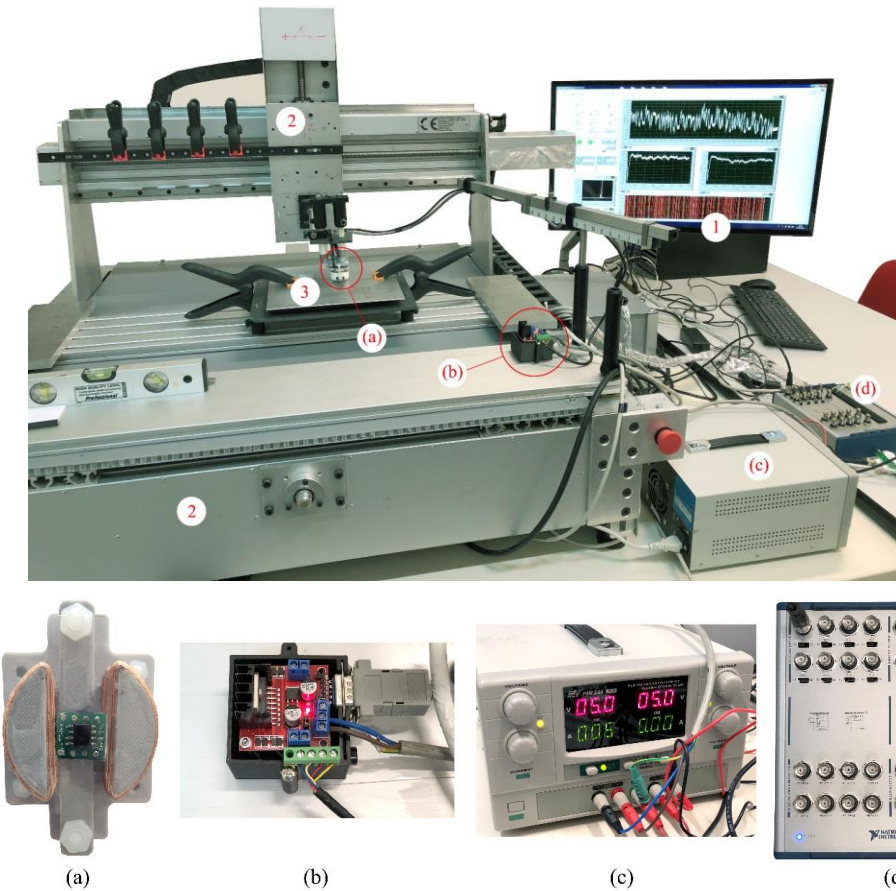

Figure 6. Setup 1: (1) PC, (2) *x-y-z* Scanning system, (3) SUT, (a) EC probe, (b) L298N driver card, (c) Power supply, (d) NI-USB 6361 Multifunction I/O device

### II.4. Measurement Setup 2

The Setup 2 is shown in Figure 7. A stable 12 V DC power supply, that could be replaced by a battery, was used for power the L298N driver. All the remaining parts of the system were powered through a USB port connected to an Analog Discovery 2 (AD2), which features a two-channel arbitrary waveform generator, a two-channel 14-bit ADC, and sixteen digital I/O lines. This device -connected to a PC- provided both the digital input lines of the L298N and analog channel to measure the output voltage of the TMR2905 sensor. The needed power for the TMR sensor was provided by the V+ Power Supply pin of the AD2 board, which can output a 5 V DC signal.

As before, six digital lines were used to control the L298N driver. The connections between the AD2 and the other system components were realised through the auxiliary BNC Adapter Board from Analog Discovery. As for the Setup 1, a LabView™ VI managed both the signal generation and acquisition. With respect to the previous case, the AD2 has a smaller buffer memory than NI-USB 6361 and a lower ADC. Precisely, the AD2 version used can control the digital lines and simultaneously acquire at maximum 4096 samples for a single-shot measurement, which is however more than enough in most of the cases for PEC and it is also enough to exploit PN excitation and PuC.

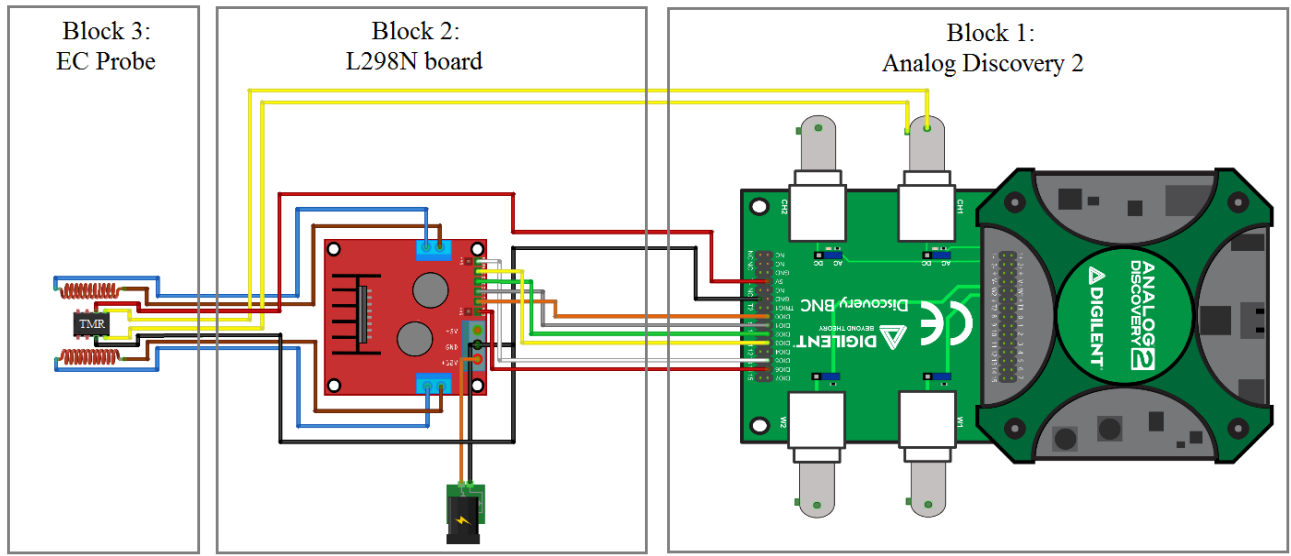

**Figure 7. Setup 2 block diagram**

This reduced acquisition length and ADC resolution is counterbalanced by the reduced costs and size of the overall measurement system, which results in a very cheap and compact setup. Furthermore, the maximum sampling rate of the AD2 is 100 MSa/s, significantly larger than the NI-USB 6361 (2 MSa/s), so it can be used in PEC applications where high frequencies up to the MHz range are needed, *e.g.* to detect surface defects, to measure thin conductive layers or insulating coatings.

### III. PSEUDO-NOISE PULSED EDDY CURRENT TESTING (PN-PEC)

#### III.1. Binary pseudo-noise sequences

Binary PN sequences are periodic discrete-time signals of "-1's" and "1's" having spectral and randomness properties like those of white noise. Many types of codes fall within the PN group such as (i) the *m*-sequences, which are derived from the Galois field theory; (ii) the Legendre sequences, $\ell$-sequences, defined by the Legendre symbol, (iii) the Kasami and Gold codes generated from *m*-sequences, and many others. Some of them, such as the *m*-sequences, can be generated by a Linear-Feedback Shift Register (LFSR) machine. They are also called maximum length sequences (MLS) because they exhibit the longest possible period for a given *N*-tap shift register. In fact, *m*-sequences exist only for period lengths $L = 2^N - 1$. Differently, $\ell$-sequences exist for any period length $L = p > 2$, where $p$ is a prime number. This is because a $\ell_p[k]$ sequence is defined from the Legendre symbol $\left(\frac{k}{p}\right)$ as for Eq. (2) [22] [23].

$$\ell_p[k] = \left(\frac{k}{p}\right) \equiv \begin{cases} 0 & \text{if } k \equiv 0 \ (\text{mod } p) \\ 1 & \text{if } k \text{ is a quadratic residue (mod p)} \\ -1 & \text{if } k \text{ is a non-quadratic residue (mod p)} \end{cases} \quad (2)$$

Despite the difference in the available lengths, and the "0" in Eq. (2) that makes $\ell$-sequences "almost" binary, *m*- and $\ell$-sequences have very similar numerical properties so, all the following considerations/algorithms apply to both in the same way. However, in the experiments reported here, we used $\ell$-sequences as they allow a more flexible tuning of the length.

PN sequences are fruitfully exploited in different applications such as telecommunication [24], encryption [25], acoustics [26], medical diagnostics [27]. In some of them, *e.g.* encryption, the numerical sequences are exploited, in other ones they are used for generating analog waveforms, where at each bit is associated a positive or negative value held for a characteristic bit duration $T_{bit}$. This is the case of NDT applications, and of the present paper. Precisely, starting from the sequence $\ell_p[k]$ and the value of $T_{bit}$, the analog waveform $s_p(t, T_{bit})$ is defined as:

$$s_p(t, T_{bit}) = \sum_{k=0}^{L-1} A\, \ell_p[k]\{\theta(t - kT_{bit}) - \theta(t - (k+1)T_{bit})\} \quad (3)$$

Note that the waveform amplitude *A* is the same of the standard PEC excitation defined in Eq. (1). What is of interest in these waveforms, in both their analog and digital representations, is their spectrum and correlation function that make them suitable for ECT applications. Figure 8 reports the examples of the sequence $\ell_{19}[k]$, the generic analog waveform $s_{19}(t, T_{bit})$ and the relative spectra defined for $T_{bit} = 50\ \mu s, 100\ \mu s$. Apart for the DC term, which do not generate eddy current, $s_p(t, T_{bit})$ has the same spectrum shape of a rectangular pulse $\Pi(t, T_{bit})$, see the inset in Figure 1. By calling $f_{bit} = \frac{1}{T_{bit}}$ the update rate of the waveform, then $BW \cong \frac{f_{bit}}{2}$.

This is of practical utility if frequency domain analysis must be implemented: one can replace the short pulse of Eq. (1) with an arbitrarily long sequence of the same amplitude having the same spectrum but with *L*-times larger excitation energy. Moreover, unlike averages, there is no dead time in the measurement, as the excitation is always active during all the measurement process. This implies that, if the PN waveform is sufficiently long, precisely equal or longer than the single pulse output, the SNR is saturated [28].

In addition, the next subsection will show how it is possible to estimate the response of the system to a $T_{bit}$-long rectangular pulse $\Pi(t, T_{bit})$, from the PN waveform output using PuC process, to retrieve the output of a standard PEC measurement.

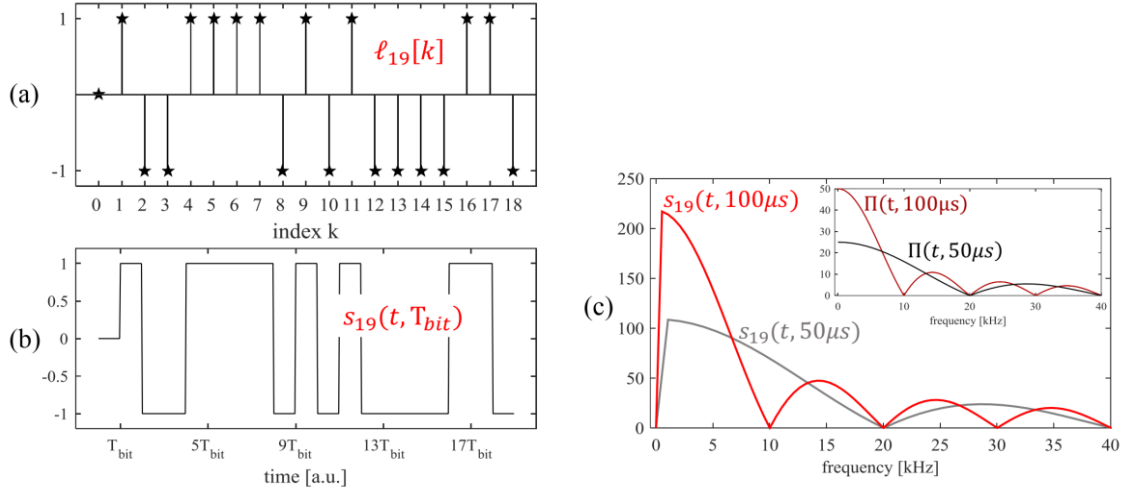

**Figure 8.** (a) Legendre sequence $\ell_p[n]$ for L=19 and A=1, (b) corresponding analog waveform $s_{19}(t, T_{bit})$, (c) spectrum amplitude of $s_{19}(t, T_{bit})$ for $T_{bit} = 50, 100$ µs. In the inset the spectrum amplitude of $\Pi(t, T_{bit})$, for $T_{bit} = 50, 100$ µs is reported

### III.2. Pulse-Compression with pseudo-noise sequences

Pulse compression (PuC) is a correlation-based technique developed for estimating the impulse response $h(t)$ of a linear time-invariant (LTI) system in noisy environments and, more in general, when high SNR values are aimed. In practice, the short excitation standardly used in many NDT applications (e.g. in PEC, in pulse-echo ultrasound, in pulsed thermography) is replaced by a broadband coded waveform and the response to the pulsed excitation is estimated by cross-correlating the output signal with the input one or with a signal derived from it. This is done to deliver more energy to the system to increase the measurement SNR. In the case of ECT, PuC can be applied if a non-ferromagnetic SUT is considered and non-linear phenomena are avoided, and some results were recently reported in [29] where a swept-frequency chirp excitation was used.

The basic theory at the heart of the technique is the following: suppose that two ideal signals $a(t)$ and $b(t)$ exist such that their cross-correlation $\Phi_{a,b}(t)$ is a Dirac delta function $\Phi_{a,b}(t) = \delta(t)$, then the impulse response of an LTI system can be reconstructed as follows:

1) use $a(t)$ as input of the system;
2) correlate the output signal $y(t)$ with $b(t)$;
3) thanks to the linearity, since $y(t) = [a * h](t)$, where $*$ denotes the convolution, by correlating $y(t)$ with $b(t)$ one obtains: $\Phi_{y,b}(t) = [\Phi_{a,b} * h](t) = h(t)$.

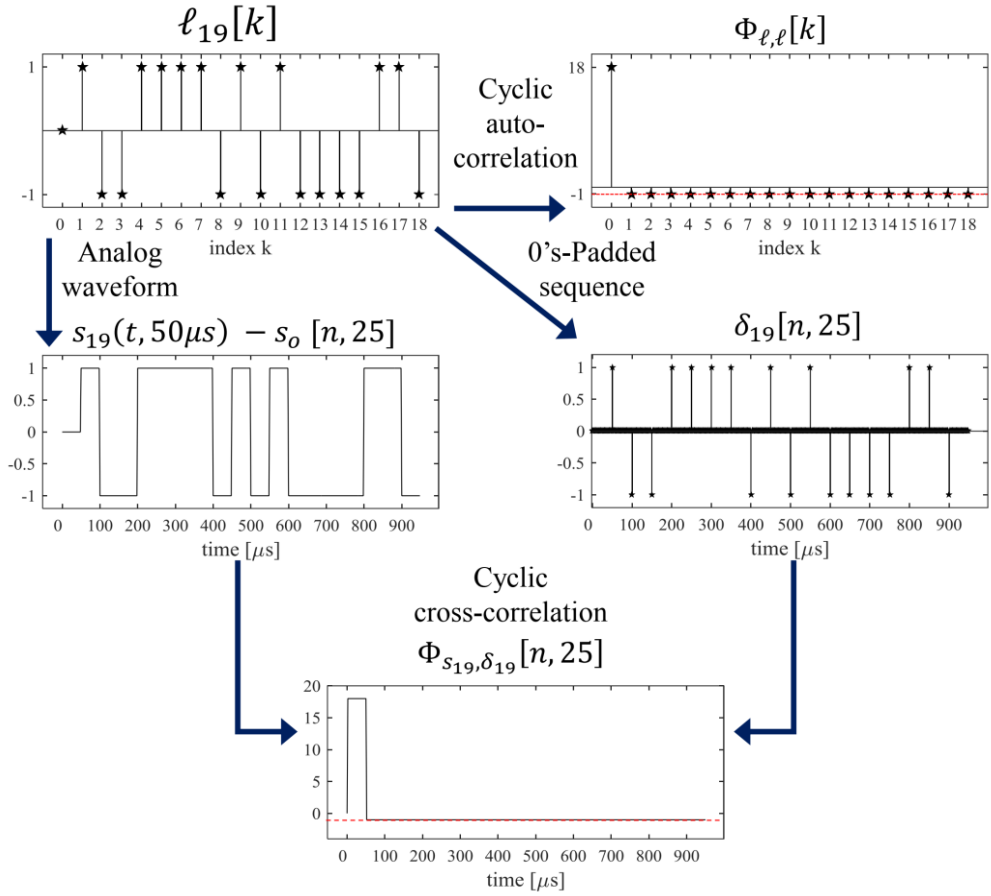

**Figure 9.** Schematic representation of the cyclic cross-correlation of the $\ell$-sequence of length 19

Unfortunately, an ideal cross-correlation does not exist for any pair of (even long) signals with finite length. Further, in most of the applications, digital signals are obtained at the end of the measurement process, so the basic PuC theory must be tailored case-by-case to approximate as much as possible the ideal condition. However, by using the PN-PEC scheme here proposed, the response $h_{PEC}(t) = [\Pi(t,T) * h](t)$ of the system to a rectangular pulse $\Pi(t,T)$, can be estimated with high accuracy.

To do this, the PuC algorithm described in [30] is used, which exploits the cyclic-correlation property of the PN sequences and of the waveforms $s_p(t, T_{bit})$. The reader is referred to that article for a comprehensive theoretical analysis of the many benefits of PuC, here the key points are reported detailing step-by-step the measurement procedure implemented.

To understand how the procedure works, it is worth stressing these aspects:

- the output signal $y(t)$ from the TMR is sampled in time at fixed sampling rate $f_S$ ($y(t) \rightarrow y[t = n\Delta t] = y\left[t=\frac{n}{f_S}\right] \equiv y[n]$), thus also the PEC response estimate is a discrete time signal with the same sampling rate $h_{PEC}\left[t = \frac{n}{f_S}\right] \equiv h_{PEC}[n]$. $f_S$ must be chosen such that $T_{bit} = \Delta t \times n_{s_{bit}}$ is a multiple of the sampling interval, hence $f_S = n_{s_{bit}} f_{bit}$;

- it is useful to introduce two discrete time signals (see Figure 9) with the same duration of $n_{s\,bit} \times L = N_0$ samples: $s_p[n, n_{s\,bit}]$, which results from sampling $s_p(t, T_{bit})$ at $f_S$, and $\delta_p[n, n_{s\,bit}]$. The former is defined by $s_p[n, n_{s\,bit}] = A\ell_p\lfloor n/n_{s\,bit} \rfloor$, where $\lfloor \cdot \rfloor$ is the floor operation, the latter is a sequence of unit pulses with amplitude defined by the $\ell_p[k]$ values:

$$\delta_p[n, n_{s\,bit}] = \begin{cases} \ell_p\left\lfloor \dfrac{n}{n_{s\,bit}} \right\rfloor & \text{if } n \equiv 0 \ (mod\ n_{s\,bit}) \\ 0 & \text{elsewhere} \end{cases} \quad (4)$$

- the correlation between two periodic discrete time signals a[n] and b[n] having the same period of N samples is periodic as well. The single period is called the cyclic correlation $\Phi_{a,b}[n]$. According to the convolution theorem for discrete time signals, $\Phi_{a,b}[n]$ is expressed by:

$$\Phi_{a,b}[n_{mod\ N}] = ifft\big(fft(a_0[n]) \cdot conj(fft(b_0[n]))\big) = ifft(fft(a_0[n]) \cdot fft(b_0[-n])) \quad (5)$$

where $a_0[n]$ and $b_0[n]$ represent just a single period of the sequences, $fft$ and $ifft$ stand for the direct and the inverse discrete Fourier transform, $b_0[-n]$ is the time-reversed replica of $b_0[n]$;

- PN-codes such as $m$- or $\ell$-sequences of length $L$ have an almost ideal cyclic-autocorrelation function [31, 30]:

$$\Phi_{l,l}[k] = \begin{cases} L - 1 & \text{if } k = 0 \ (mod\ L) \\ -1 & \text{elsewhere} \end{cases} = L\delta[k] - 1 \quad (6)$$

where $\delta[k]$ is the unit pulse (see Figure 9).

- a relation similar to that of Eq. (6) is obtained from the cyclic cross-correlation between $s_p[n, n_{s\,bit}]$ and $\delta_p[n, n_{s\,bit}]$:

$$\Phi_{s_p,\delta_p}[n_{s\,bit}, n_{s\,bit}] = \begin{cases} A(L-1) & \text{if } 0 \leq n < n_{s\,bit} \\ -A & \text{elsewhere} \end{cases} = A(L\Pi[n, n_{s\,bit}] - 1) \quad (7)$$

where $\Pi[n, n_{s\,bit}]$ is the digital version of $\Pi(t, T_{bit})$ at the sampling rate $f_s$ (see Figure 9).

Starting from these points above, the following PuC procedure is defined that uses a $s_p(t, T_{bit})$ waveform as excitation signal. To get a deeper insight of the process, Figure 2 reports the main signals introduced above, while Figure 3 summarizes the measurement procedure.

1) a $T_{bit}$ value is chosen to define the excitation bandwidth: $BW \cong \dfrac{1}{2T_{bit}}$;
2) an $L$-bit long PN sequence $\ell_L[k]$ is selected to reach an excitation duration equal to $T_0 = T_{bit} \times L > T_{PEC}$, where $T_{PEC}$ is the expected duration of $h_{PEC}(t)$, the system output to a rectangular excitation pulse $\Pi(t, T_{bit})$;

3) a periodic analog waveform $s(t)$ is generated starting from the fundamental period $s_p(t, T_{bit})$ defined in Eq. (2) (actually just two periods are enough);
4) the SUT is excited by $s(t)$ and its output response $y(t)$ is measured and sampled at sampling frequency $f_s$;
5) the digital signal $y[n]$ is periodic with period $N_0$ starting from $n > N_0$ (the transient $y_{tr}[n]$ certainly extinguishes for $n > N_0$);
6) the steady-state response $y_0[n]$ (i.e. the second period) is processed to obtain the PEC response estimate:

$$\hat{h}_{PEC}[n] = \Phi_{y_0,\delta_p}[n] = \{\Phi_{s_p,\delta_p} * h\}[n] = A\{(L\Pi[n, n_{s_{bit}}] - 1) * h\}[n] \cong Lh_{PEC}[n] \tag{8}$$

which is the key result of PuC.

The entire process described above is depicted in Figure 10.

The single pulse response is thus estimated as if the system was being excited by a signal L-times larger in amplitude, hence significantly reducing the effect of any additive noise. Moreover, the constant term "−1" in the previous Eq. (6) can be removed as explained in [30].

It is also possible to apply PEC standard analysis to PuC data. For instance, in [29], the lift-off invariant point was found after PuC.

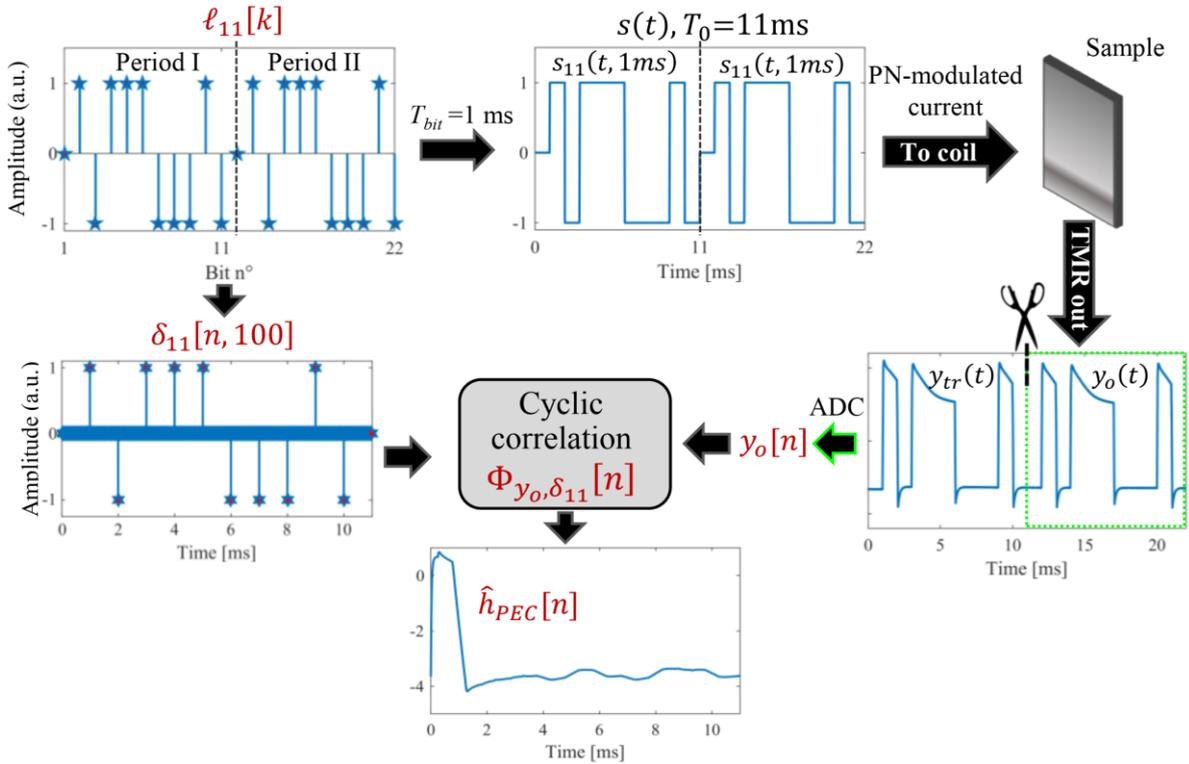

Figure 10. PN-PEC measurement procedure: starting from $\ell_p[n]$ and $T_{bit}$, the two-period long excitation signal is generated. The TMR output is sampled and the second period is processed to retrieve the estimate of a PEC standard test.

## IV. EXPERIMENTAL RESULTS

### IV.1. Benchmark sample

The experimental data were collected exploiting the PN-PEC procedure on a specimen having known artificial defects. This was a plate made of 2024-T3 aluminium alloy with electric conductivity equal to 18.8 MS/m, magnetic permeability of 1.26 µH/m, and thickness of 2 mm.

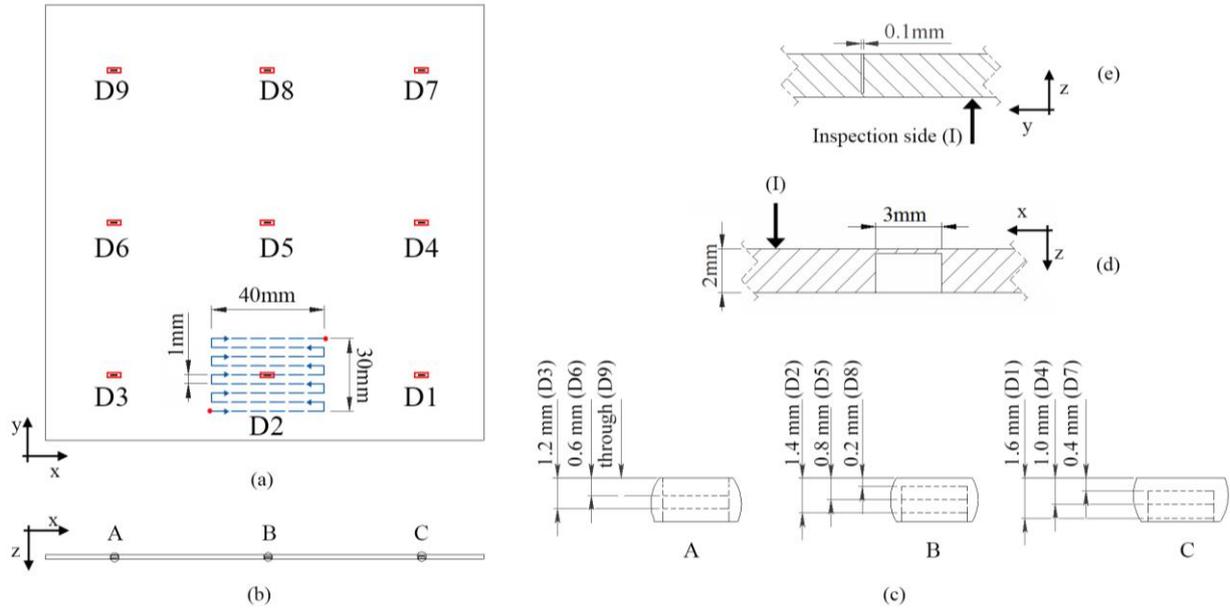

Figure 11. (a) Top view of the sample with nine defects and a sample scan grid path around defect D2; (b) side view of the sample (c); detailed views of A, B and C; (d) side view of the defect D8 (e); front view of defect D8.

All the defects were small notches having a length of 3 mm and width of 0.1 mm, with varying depths from the inspection surface. Starting from $D_1$ to $D_8$ their depth varied respectively from 1.6 mm to 0.2 mm, with a step of 0.2 mm. $D_9$ is a through-thickness slit. For each defect except $D_9$, a set of measurements were collected over a regular grid of $[N_x = 80 \times N_y = 60]$ measurement points with spatial resolution of 0.5 mm × 0.5 mm. A sketch of the sample and the typical scan trajectory over a defect are depicted in Figure 11.

**Table 2. Subsurface and surface defect depths [mm]**

| Defect No. Location | $D_1$ | $D_2$ | $D_3$ | $D_4$ | $D_5$ | $D_6$ | $D_7$ | $D_8$ | $D_9$ |
|---|---|---|---|---|---|---|---|---|---|
| Surface | 0.40 | 0.60 | 0.80 | 1.00 | 1.20 | 1.40 | 1.60 | 1.80 | 2.0 |
| Subsurface | 1.60 | 1.40 | 1.20 | 1.00 | 0.80 | 0.60 | 0.40 | 0.20 | 0.0 |

Further, a 2 mm thick sound plate was subsequently placed above the sample, with the configuration shown in Figure 12, and other set of scans were acquired.

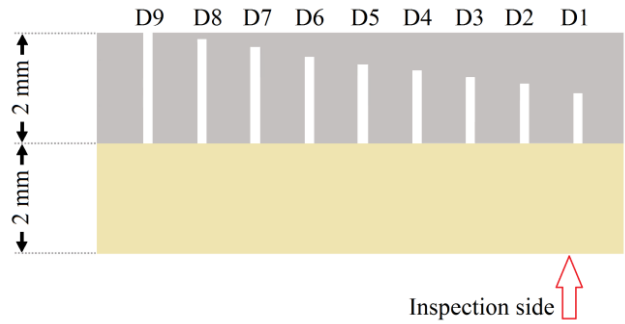

Figure 12. Sample configuration 2 consisting of two layers of aluminium

## IV.2. PN-PEC Results

The results reported in this section aim at demonstrating the effectiveness of the measurement system and the main features of the PN-PEC approach: (1) at the end of the PN-PEC procedure signals are obtained that are indistinguishable from those obtained by standard PEC but (2) a high SNR is achieved allowing very small defects to be detected and imaged.

Related to the point (1), it will be also shown how PN-PEC output exhibits LOI property. All the figures reported have been produced by processing data collected with the Setup 1 and using the parameters $L = 67$ and $T_{bit} = 100$ μs to ensure a high SNR for detecting all the defects tested. However, many tests were collected with smaller $L$ and $T_{bit}$ values and with the Setup 2 giving similar results. SNR increases linearly with $L$, as demonstrated in [30]. The scans were acquired at various lift-off values ranging from 0 mm to 3 mm with a step of 1 mm. The two configurations (a) and (b) of the excitation magnetic field were both tested (see Figure 4).

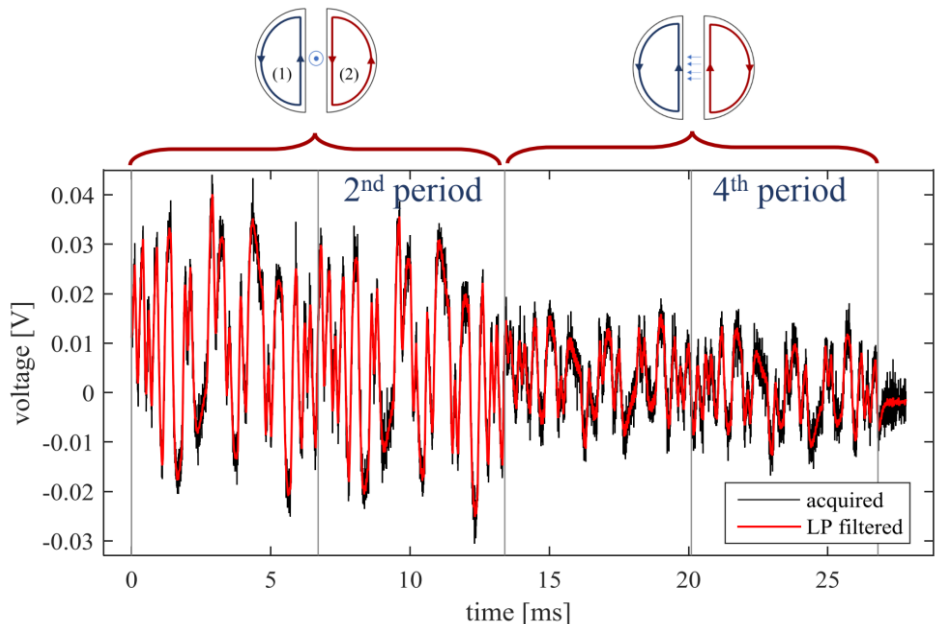

Figure 13. An example of the acquired TMR output signal before (black) and after low-pass filtering (red). The two excitation configurations were used consecutively and, for everyone, the 2nd period, corresponding to the steady state response, is processed with the PuC algorithm.

To do this while saving measurement time, a unique digital pattern was created by appending four repetitions of the PN basic waveform $s_{67}(t, 100\ \mu s)$ of duration $T_0 = 6.7$ ms. To ensure firstly the perpendicular field excitation and then the tangential one, the coils (1) and (2) were driven by the current signals $I_{(1)} = [+s_{67}, +s_{67}, +s_{67}, +s_{67}]$ and $I_{(2)} = [+s_{67}, +s_{67}, -s_{67}, -s_{67}]$, where $s_{67}$ stands for $s_{67}(t, 100\ \mu s)$. This is a further demonstration of the flexibility of the excitation system here adopted.

By assuming $T_0 > T_{PEC}$, the second and the fourth periods of the TMR output signal correspond to the steady-state response of the two configurations and they are then independently processed by following the PuC procedure described above. Figure 13 shows a typical TMR signal acquired over a sound point of the sample. The switch from one configuration to the other is evident. There is also clearly visible a high-frequency noise so a low-pass filter with $f_c = 50$ kHz is applied in post-processing to remove high-frequency noise before PuC. Since Setup 1 has a large buffer size, the whole response was acquired as a unique waveform, however the 2$^{nd}$ and the 4$^{th}$ periods can be acquired independently thus reducing of a factor 4 the memory buffer required or a single shot.

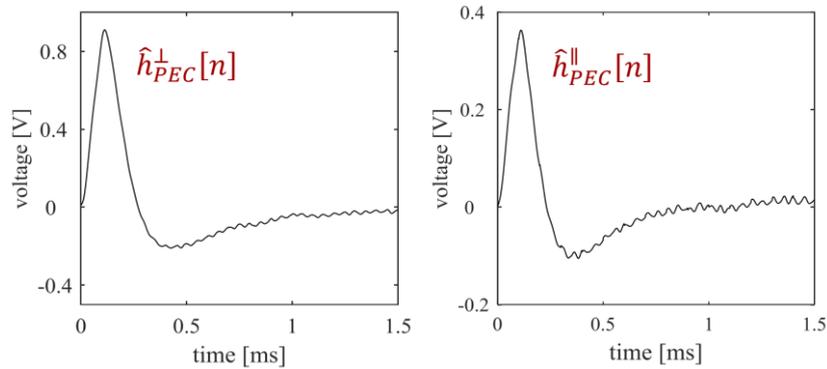

**Figure 14. Example of PN-PEC output signals $\hat{h}_{PEC}^{\perp}[n]$ and $\hat{h}_{PEC}^{\parallel}[n]$ collected on a sound point and corresponding to the perpendicular and tangential field excitation respectively.**

Figure 14 is reported as an example of the two signals $\hat{h}_{PEC}^{\perp}[n]$ and $\hat{h}_{PEC}^{\parallel}[n]$ reconstructed after PuC and corresponding to the perpendicular and tangential field respectively. The PN-PEC outputs have the typical behaviour of a PEC output signal, as expected.

PuC was then applied to all the measurements collected in the various scans, and time and frequency features were extracted through the same process introduced in [29] and imaging protocols based on these features were implemented. It was found that the configuration (b) with the primary magnetic field lines parallel to the sample ensured the highest sensitivity to defect with the existing sensor. As an example of this, Figure 15 (top) reports two *x-y* images (C-scans) representing the value of $\hat{h}_{PEC}^{\perp}[n]$ and $\hat{h}_{PEC}^{\parallel}[n]$ at $t = 0.5$ ms collected over the defect D5 at $LO = 0$ mm. The typical quadrupolar defect pattern [29] has a higher contrast for the $\hat{h}_{PEC}^{\parallel}$ case.

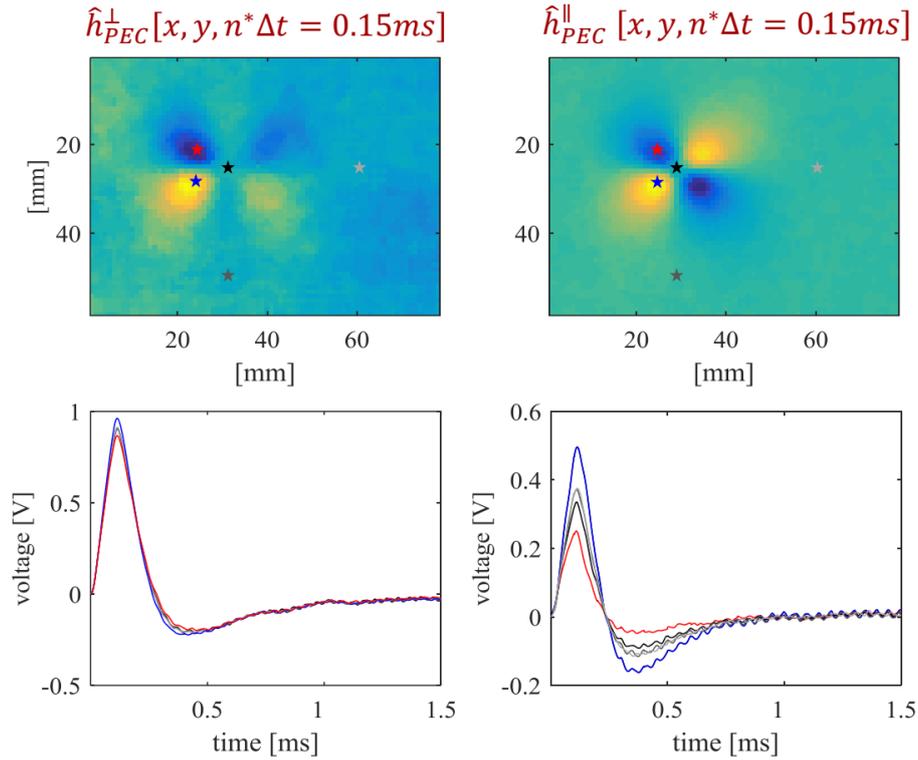

Figure 15. Example of $\hat{h}^{\perp}_{PEC}[n]$ and $\hat{h}^{\parallel}_{PEC}[n]$ signals reconstructed after the PuC and corresponding to the perpendicular and tangential field respectively. Top: C-scan image at 0.5 ms, Bottom: A-scan signal corresponding to the marked points in C-scan

To further corroborate this result, in Figure 15 (bottom) some $\hat{h}^{\perp}_{PEC}[n]$ and $\hat{h}^{\parallel}_{PEC}[n]$ curves are reported, which correspond to the marked points over the defect signature or on sound areas. While all the $\hat{h}^{\perp}_{PEC}[n]$ curves are very close to each other, in the case of $\hat{h}^{\parallel}_{PEC}[n]$ the signals corresponding to the defect differentiate significantly from those acquired on sound points.

This demonstrates the higher sensitivity of $\hat{h}^{\parallel}_{PEC}[n]$, so the results shown hereinafter refer only to this case and will complete the overview of the potentialities of the PN-PEC approach.

### IV.3. Lift-off invariant point

Since the PN-PEC output $\hat{h}_{PEC}[n]$ is in fact indistinguishable from the output of a PEC experiment, the LOI point/region should be found on $\hat{h}_{PEC}[n]$. To verify this, in Figure 16 (top) the $\hat{h}^{\parallel}_{PEC}[n]$ curves obtained on a sound point at LO=0,1,2,3 mm are plotted and a zoomed view is also reported showing the presence of a LOI point around $t = 260\ \mu s$.

The presence of the LOI point was confirmed by performing imaging in the time domain in presence of LO variation. The method presented in [29] was adopted and images were reconstructed for a LO that varies randomly point-by-point with uniform probability between the values 0 mm to 3 mm. Despite of the harsh, perhaps extreme, LO noise simulated, for $t = 260\ \mu s$, the defect pattern can be still distinguished, although noisy. For an even small change in time, the pattern becomes barely or not at all visible.

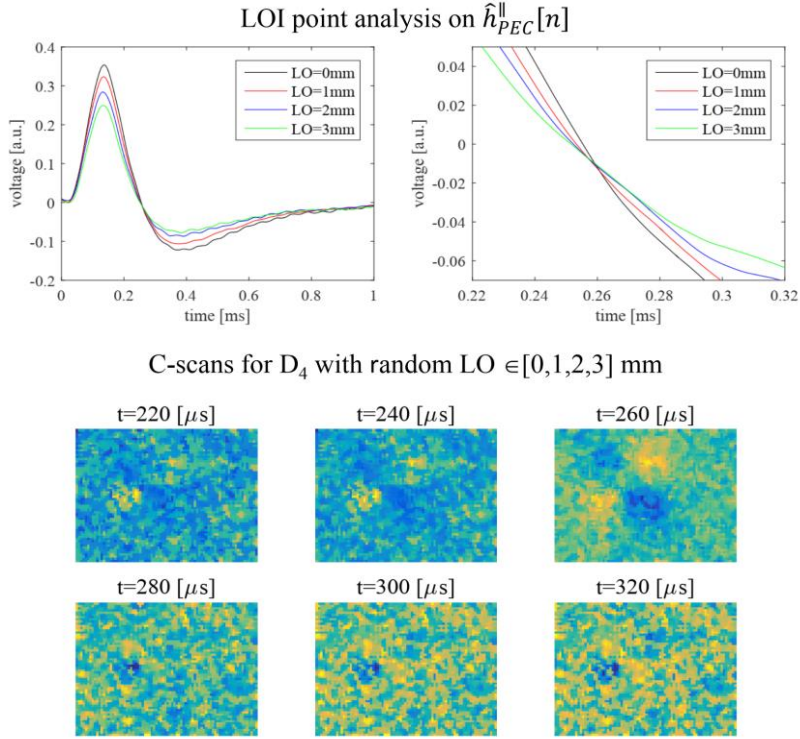

Figure 16. Simulated random lift-off case and retrieved LOI point reported for defect D4

### IV.4. Defect detection capability

The defect detection capability of the system is evaluated in Figure 17 and Figure 18. Figure 17 reports the best C-scans obtained by imaging the time-amplitude of $\hat{h}_{PEC}^{\parallel}[n]$ at different time values for the defects D1-D8 and for all the LO values. All the defect patterns are visible for LO = 0,1 mm, while for LO=2, 3 mm deeper defects become barely visible. It is interesting also to see the spread of the defect patterns as the LO increases. Figure 17reports the best C-scans obtained for the sample configuration illustrated in Figure 12 and for LO=0 mm only.

## V. CONCLUSION AND FUTURE WORK

As discussed through this paper, there were many efforts devoted to optimization and improvement of the eddy current devices. This work is a breakthrough in new generation of low-cost handy instruments with high efficiency in terms of defect detection capability while overcoming the classic problem of lift-off in eddy current testing. The system introduced in this work uses the powerful characteristics of both pseudo-noise sequences and the pulse-compression in order to improve the signal-to-noise ratio and achieve comparable results with respect to standard pulsed eddy current.

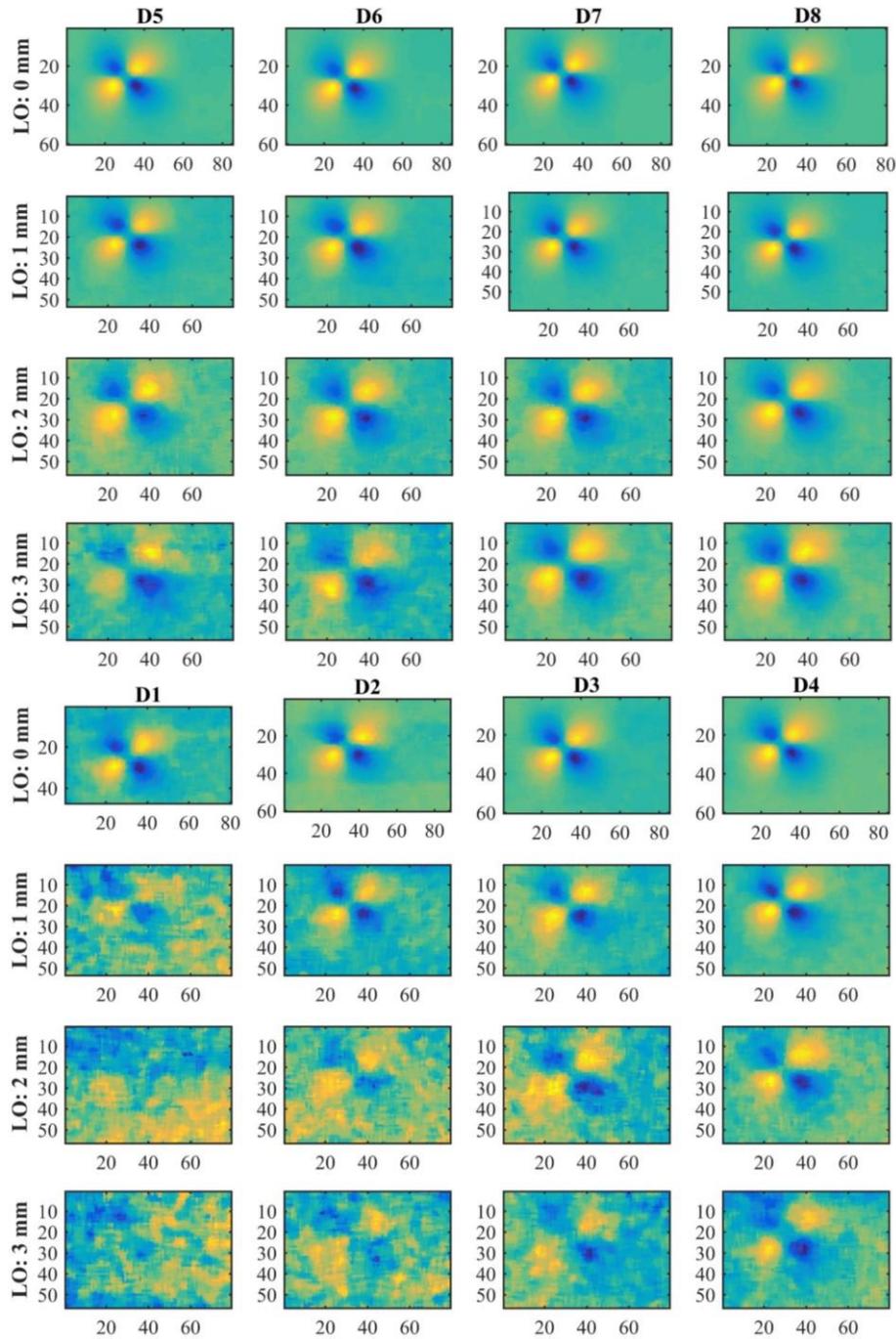

**Figure 17. Best C-scan image slices for defects D1-D8 and all lift-off cases for one-layer sample configuration**

The so-realised instrument ensures a high defect detection capability and can be easily replicated using a variety of similar components, and further improved exploiting single-board PC such as Raspberry Pi, Arduino, NVIDIA Jetson Nano, etc.

This work was the first step toward development of an integrated compact and reliable eddy current system being able to be used in industrial applications. However, future work shall focus on design of the excitation board to be more compatible with this specific application, an integrated battery-operated power supply, an integrated software with a small processing system such as

Raspberry Pi to enable mobile data acquisition and processing. Also, the scanning process can be optimized by using an encoder or motion sensor integrated into the eddy current probe.

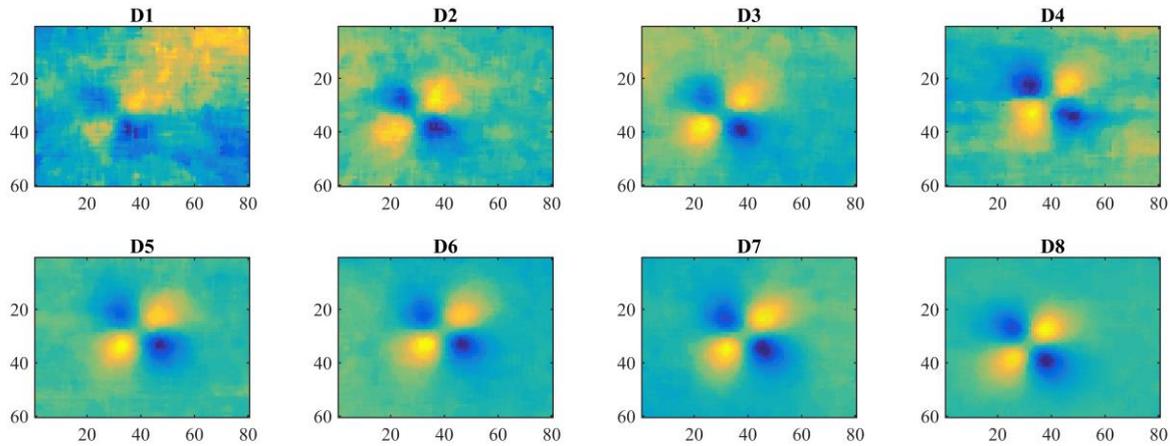

Figure 18. Best C-scan image slices for the two-layer configuration at LO=0 mm and for all defects

ACKNOWLEDGEMENT

This research work has been partially supported from the European Union's Horizon 2020 research and innovation programme under the Marie Skłodowska-Curie grant agreement No 722134 – NDTonAIR.